\documentclass[aps,twocolumn]{revtex4}
\usepackage{amsmath}
\usepackage{amsfonts}
\usepackage{mathrsfs}
\usepackage{graphicx}
\usepackage{times}
\usepackage{color}
\usepackage{multirow}
\begin{document}

\title{Chiral odd Chern number lattice supersolidity with tunable unpaired Majorana fermions in a Rydberg-dressed Fermi gas}

\author{Shuai Li}
\affiliation{Ministry of Education Key Laboratory for Nonequilibrium Synthesis and Modulation of Condensed Matter,Shaanxi Province Key Laboratory of Quantum Information and Quantum Optoelectronic Devices, School of Physics, Xi'an Jiaotong University, Xi'an 710049, China}

\author{Rui Tian}
\affiliation{Ministry of Education Key Laboratory for Nonequilibrium Synthesis and Modulation of Condensed Matter,Shaanxi Province Key Laboratory of Quantum Information and Quantum Optoelectronic Devices, School of Physics, Xi'an Jiaotong University, Xi'an 710049, China}

\author{Min Liu}
\affiliation{Ministry of Education Key Laboratory for Nonequilibrium Synthesis and Modulation of Condensed Matter,Shaanxi Province Key Laboratory of Quantum Information and Quantum Optoelectronic Devices, School of Physics, Xi'an Jiaotong University, Xi'an 710049, China}

\author{Maksims Arzamasovs}
\affiliation{Ministry of Education Key Laboratory for Nonequilibrium Synthesis and Modulation of Condensed Matter,Shaanxi Province Key Laboratory of Quantum Information and Quantum Optoelectronic Devices, School of Physics, Xi'an Jiaotong University, Xi'an 710049, China}

\author{Bo Liu}
\email{liubophy@gmail.com}
\affiliation{Ministry of Education Key Laboratory for Nonequilibrium Synthesis and Modulation of Condensed Matter,Shaanxi Province Key Laboratory of Quantum Information and Quantum Optoelectronic Devices, School of Physics, Xi'an Jiaotong University, Xi'an 710049, China}

\begin{abstract}
There is growing interest to search the chiral Majorana fermions that could arise as the quasi-particle edge state of a two-dimensional topological state of matter. Here we propose a new platform, i.e., a two-dimensional chiral odd Chern number lattice supersolid state, for supporting multiple number-tunable chiral Majorana fermions from a single component Rydberg-dressed Fermi gas in an optical lattice. The attractiveness of our idea rests on the fact that by introducing the unveiled competition between two distinct length scales, i.e., lattice period and the distance of resonant Rydberg-dressing, can provide a new way to manipulate the spatial dependence of both the strength and sign of the effective Rydberg-dressed interaction. Such a designed effective interaction turns out, can induce an unveiled odd Chern number lattice supersolid state, which is confirmed by both the mean-field and Monte Carlo
calculations. Furthermore, we also find that the spontaneously formed density modulation resulted from the discrete translational symmetry breaking provides a natural way of tuning the system's topology arising from the
superfluidity induced by the $U(1)$ symmetry breaking. It thus provide an alternative way for manipulating
the chiral Majorana fermions, which would be useful in topological quantum computation.

\end{abstract}

\maketitle
Pursuit of chiral Majorana fermions (CMFs) has attracted intensive interests in recent years ~\cite{nayak2008s}
. The non-Abelian braiding of CMFs is considered as the basic building block for fault tolerant topological quantum computations ~\cite{karzig2017scalable,lian2018topological,zhou2019non}. So far, several systems were proposed to realize CMFs. One example of
hosting the chiral Majorana fermion mode (CMFM) is the 2D topological superconductor, like the $p_x+ip_y$ superconductivity in the liquid $^{3}$He ~\cite{volovik2003universe} and strontium ruthenates ~\cite{kallin2012chiral}, which are in the same universality class as fractional quantum Hall states~\cite{read2000paired}. However,
the fate of 2D topological superconductivity in electronic matter remains debatable. In the field of ultracold atoms, this phase was predicted to appear via manipulating $p$-wave interactions (or equivalent one), such as utilizing the $p$-wave Feshbach resonance, artificial spin-orbit coupling or dipolar interactions ~\cite{gurarie2007resonantly, regal2003tuning, galitski2013spin, baranov2012condensed}. But, the experimental challenges in the above proposals, such as three-body loss, heating problem or ultracold chemical reactions, are still desired to future breakthroughs. Another approach proposed to get around is to hybridize materials of topological and superconducting
properties, e.g., semiconductor-superconductor heterostructures, a helical magnetic structure on top of superconductors and topological insulators coupled with superconductors ~\cite{sau2010generic, alicea2010majorana, nadj2013proposal, braunecker2013interplay, fu2008superconducting, akhmerov2009electrically, qi2010chiral}. This approach nevertheless requires advanced material engineering.

Here we report the discovery of a new many-body phase, i.e., a chiral odd Chern number lattice supersolid (CLSS) state, which can support the number-tunable CMFM. Distinct from topological superconductors, in our proposed CLSS state, not only is the $U(1)$ symmetry broken, but the discrete translational symmetry is also broken. More interestingly, it is shown that such a discrete translational symmetry breaking induced density modulation provides a new tool, which is missing in topological superconductors, for manipulating the topological nature of CLSS and thus to support the number-tunable CMFM. We shall introduce this with a specific model of Rydberg-dressed Fermi atoms in an optical lattice, to be described below. Recently, the research of Rydberg atoms and Rydberg-dressed atoms has evolved rapidly ~\cite{weimer2010rydberg, lukin2001dipole, saffman2010quantum, browaeys2016interacting, karpiuk2015imaging}, where an effective Rydberg-dressed interaction (RDI) shows high controllability and thus haven been recognized for their potential in quantum simulation and quantum information ~\cite{schauss2015crystallization, zeiher2016many, hollerith2019quantum, zeiher2017coherent, jau2016entangling, guardado2018probing, borish2020transverse}. Lots of interesting many-body phases induced by the RDI, such as a supersolid droplet phase, a bright soliton, a topological superfluid and topological density waves, have been predicted ~\cite{henkel2010three, henkel2012supersolid, buchler2007strongly, cinti2010supersolid, pupillo2010strongly, maucher2011rydberg, xiong2014topological, li2015exotic}. Distinct from previous studies, the new idea here is to
utilize the competition between two different length scales, i.e., the period of an optical lattice and the distance of resonant Rydberg dressing ~\cite{ates2012dissipative, li2013nonadiabatic}, as a new tool to manipulate the RDI,
which is motivated by the recent experimental advances in Rydberg-dressed atoms in  optical lattices ~\cite{jau2016entangling, zeiher2016many}. Interestingly, it is shown that both the interaction strength and sign of RDI can be
engineered to be spatially dependent. And such a designed RDI can induce an unveiled CLSS state.

\textit{Effective model $\raisebox{0.01mm}{---}$} Let us consider a single-species Fermi gas held in a 2D square optical lattice, where atoms are coupled to their Rydberg states through the double Rydberg dressing scheme ~\cite{galitski2013spin, schauss2015crystallization, zeiher2016many, hollerith2019quantum, zeiher2017coherent, jau2016entangling} to generate an effective RDI.  Here the ground state atom is simultaneously coupled to two Rydberg states by applying one blue-detuned and one red-detuned lasers together~\cite{ lukin2001dipole, saffman2010quantum, browaeys2016interacting, karpiuk2015imaging}.  Through tuning the Rabi frequency and detuning of the off-resonant light,  the RDI between dressed-state atoms can be captured by the following form ~\cite{volovik2003universe, kallin2012chiral}:
\begin{equation}
V(r)=U_{1}(r)+U_{2}(r),  \label{1}
\end{equation}%
where  $U_{j}(r)=\tilde{C}_{6}^{(j)}/(r^{6}\mp
\tilde{R}_{j})$ with $j=1,2$ describes the distinct RDI induced by the coupling to different Rydberg state $|\tilde{R} _{j}\rangle$.
$\tilde{C}_{6}^{(j)}=\tilde{R}_{j}^{6}\Omega _{j}^{4}/8\left\vert
\Delta _{j}\right\vert ^{3}$ is the interaction strength,  where the averaged soft-core radius
 $\tilde{R}_{j}=(C_{6}^{(j)}/2\left\vert \Delta
_{j}\right\vert )^{1/6}$ and $C_{6}^{(j)}>0$ denotes the van der Waals (vdW) interaction strength of
Rydberg state  $|\tilde{R} _{j}\rangle$, which is assumed to be positive in this work.  $\Omega _{j}$ and $\Delta _{j}$
stand for the corresponding Rabi frequency and detuning, respectively.
The plus and minus signs refer to
the red- and blue- detuned lasers, respectively.

When the lattice depth is large enough, the above system can be described by the following Fermi-Hubbard model in the tight-binding regime
\begin{eqnarray}
\textstyle\mathbf{H} &=&-\sum_{<i,j>}t(c_{i}^{\dag }c_{j}+h.c.)-\mu
\sum_{i}c_{i}^{\dag }c_{i}  \notag \\
&&+\frac{1}{2}\sum_{i\neq j}V_{i-j}c_{i}^{\dag }c_{j}^{\dag }c_{j}c_{i},
\end{eqnarray}%
where $t$ is the hopping amplitude describing tunneling in the 2D plane.  $i\equiv (i_{x},i_{y})$ is the site index denoting
the lattice site $\mathbf{R}_{i}\equiv (ai_{x},ai_{y})$ with $a$ being the lattice constant.  $\mu $ is the
chemical potential. The RDI  is  given by $V_{i-j}=V(\mathbf{R}_i-\mathbf{R}_j)$. The attractiveness of our idea rests on the fact that through simultaneously tuning the lattice constant,  Rabi frequency and detuning,  both the interaction strength and sign of RDI can be engineered to be  spatially dependent in 2D plane. In the double Rydberg dressing scheme, there is a critical distance $R_{res}$ determined by the relation $2\Delta_1+C_{6}^{(1)}/R^6_{res}=0$, at which Rydberg atom pairs are resonantly excited ~\cite{volovik2003universe, karzig2017scalable, lian2018topological}. At the same time, another length scale is determined by the lattice constant. Interestingly, the competition of the above two distinct length scales can result in unusual effects on the RDI. For instance, here we consider tuning the two length scales in the regime $R_{res}/\sqrt{5}<a<R_{res}/2$, where the RDI shows following unveiled features. It is found that when varying the inter-particle distance in optical lattices, the sign of RDI
becomes highly tunable, i.e., (i) when $|\mathbf{R}_i-\mathbf{R}_j|<R_{res}$, the RDI is attractive; (ii) when $|\mathbf{R}_i-\mathbf{R}_j|>R_{res}$, the RDI is repulsive (assuming $\Omega_{1}^{4}/\left\vert \Delta_{1}\right\vert ^{3}>\Omega_{2}^{4}/\left\vert \Delta_{2}\right\vert ^{3}$). Therefore, the nearest-neighbor $V_{N}$, next-nearest-neighbor $V_{NN}$ and next-next-nearest-neighbor $V_{NNN}$ interaction in Eq. (1) are attractive, while other long-range interactions are repulsive. More interestingly, it is also shown
that the longer range attraction $V_{NNN}$ and $V_{NN}$ is engineered to be stronger than the nearest-neighbor attraction $V_{N}$. Past studies have shown that when including both attractive $V_{N}$ and $V_{NN}$ interactions between lattice fermions, two kinds of many-body phases, i.e.,
the charge density wave (CDW) and superfluid (SF) phases can appear. However, the CDW phase can only survive in the limit when
$V_{NN} \ll V_{N}< 0$~\cite{capponi2015phase,Corboz_2012}. Here, surprisingly, it is shown that our designed spatially-dependent RDI not only frees up that limitation, but also
results in the  coexistence of CDW and SF and thus induces an interesting lattice-supersolid phase (SS),  which is confirmed by both mean-field and Monte Carlo studies in the following.

\begin{figure}[h]
\begin{center}
\includegraphics[width=8.4cm]{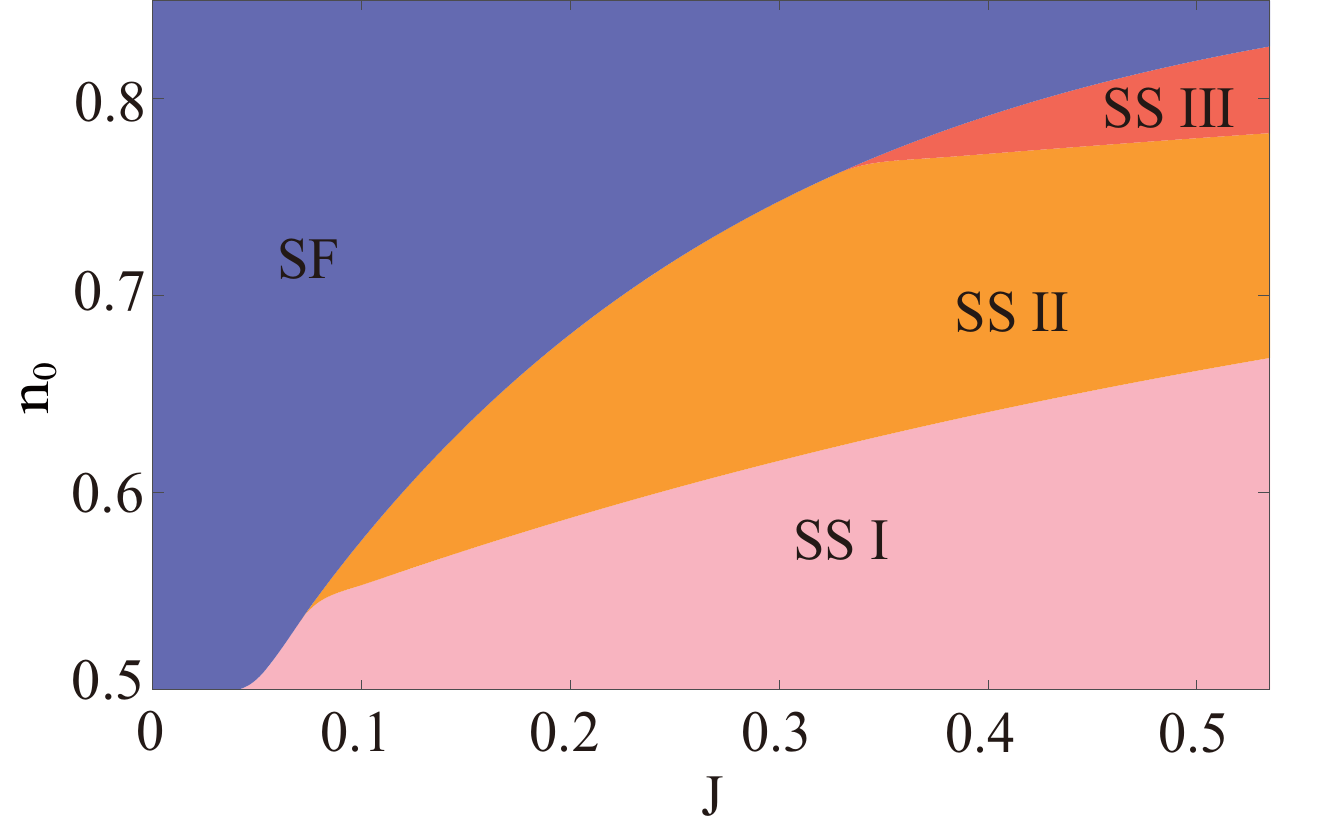}
\end{center}
\caption{(a) Zero-temperature phase diagram as a function of the average
filling $n_{0}$ and interaction strength $J$. For certain $n_{0}$, there
is a threshold of $J$. Beyond that, three topologically distinct SS phases appear.
Here $J\equiv{\tilde{C}_{6}^{(1)}}/{\tilde{R}_{1}^{6}t}$ and other parameters are
chosen as $R_{res}/a=2.08$, $\tilde{R}_{2}/\tilde{R}_{1}=1.5$, $\tilde{C}_{6}^{(1)}/\tilde{C}_{6}^{(2)}=1$.}%
\label{fig:1}%
\end{figure}

\begin{figure}[t]
\begin{center}
\includegraphics[width=8.5cm]{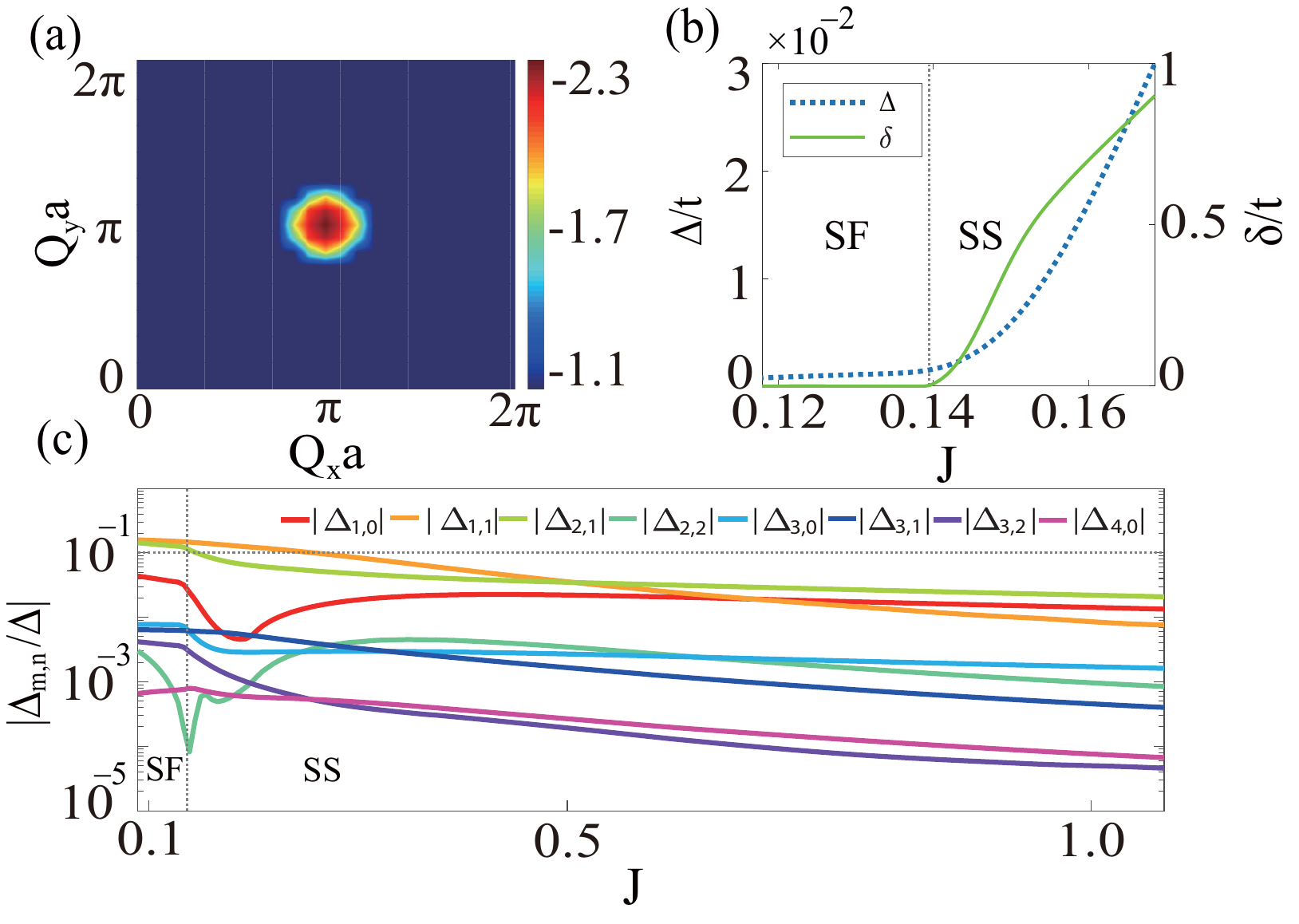}
\end{center}
\caption{(a) Mean-field energy as a function of $\mathbf{Q}$ for $J=0.18$, $n_{0}=0.6$. (b) The superfluid pairing and CDW order parameters marked by
the dashed and solid lines, respectively, where $n_{0}=0.64$. (c) The fourier series expansion of the superfluid order parameter for $n_{0}=0.64$. Other parameters
are the same as in Fig.~\ref{fig:1}.}
\label{fig:2}
\end{figure}

First, under the mean-field approximation, to describe the CDW,  we rewrite the density distribution of the system as $n_{i}=n_{0}+C\cos (\mathbf{Q}\cdot \mathbf{R}_{i})$, where $\mathbf{Q}$ represents the periodicity of density pattern and  $n_{0}=\sum \limits_{i}\langle c_{i}^{\dagger }c_{i}\rangle/N_L$ is the average filling with $N_L$ being total lattice site. Therefore, the CDW order parameter can be defined as   $\delta _{\pm \mathbf{Q}}=V(\pm \mathbf{Q})C$ with $V(\mathbf{k})=\sum_{n\neq 0}V_{n}\exp (-i\mathbf{k}\cdot \mathbf{r}_{n})$.  We also introduce the superfluid pairing order parameter as $\Delta ({\mathbf{k}})=\frac{1}{N_L}\sum_{\mathbf{k'}} V(\mathbf{k}-\mathbf{k'})\langle c_{-\mathbf{k'}}c_{\mathbf{k'}}\rangle$ and $\langle...\rangle$ stands for the expectation value in the ground state. Through minimizing the ground state mean-field energy, order parameters defined above can be obtained (see details in Supplementary
Materials (SM)). We find that there is a threshold of the interaction strength $J$ for supporting the coexistence of superfluid and CDW orders, for instance, as shown in Fig.~\ref{fig:2}(b). Regarding the CDW order, it is shown that the mean-field ground state energy is minimized at $\mathbf{Q} = (\pi/a ,\pi/a)$ (Fig.~\ref{fig:2}(a)), indicating that there is a checkerboard density pattern and the CDW order parameter can be written as $\delta\equiv\delta_{(\pi/a ,\pi/a)}$.  For the superfluidity, there is a complex superfluid order parameter with odd parity. As shown in Fig.~\ref{fig:2}(c), we apply a fourier series expansion to the superfluid order parameter, i.e., $\Delta (\mathbf{k})=\sum_{m,n}\Delta _{m,n}\sin({m}k_xa+{n}k_ya)$, and it is found that when J increasing,
the dominant component of $\Delta(\mathbf{k})$ behaves as $\Delta [\sin(2k_{x}a) + i \sin(2k_{y}a)]$, since we find that $\Delta\equiv\Delta_{2,0}=-i\Delta_{0,2}$. Because the checkerboard CDW order breaks the discrete translational symmetry and the superfluidity breaks the $U(1)$ symmetry, the coexistence of these two orders will lead a SS phase. We thus obtain the zero-temperature phase diagram as shown in Fig.~\ref{fig:1}. When fixing a certain average filling, there is a threshold of interaction strength J separating the SF and SS. Below that threshold, the ground state is a superfluid, where the CDW order vanishes. When further increasing the interaction strength, the superfluid and CDW coexist, indicating that the ground state is a SS phase.

\begin{figure}[t]
\begin{center}
\includegraphics[width=8.5cm]{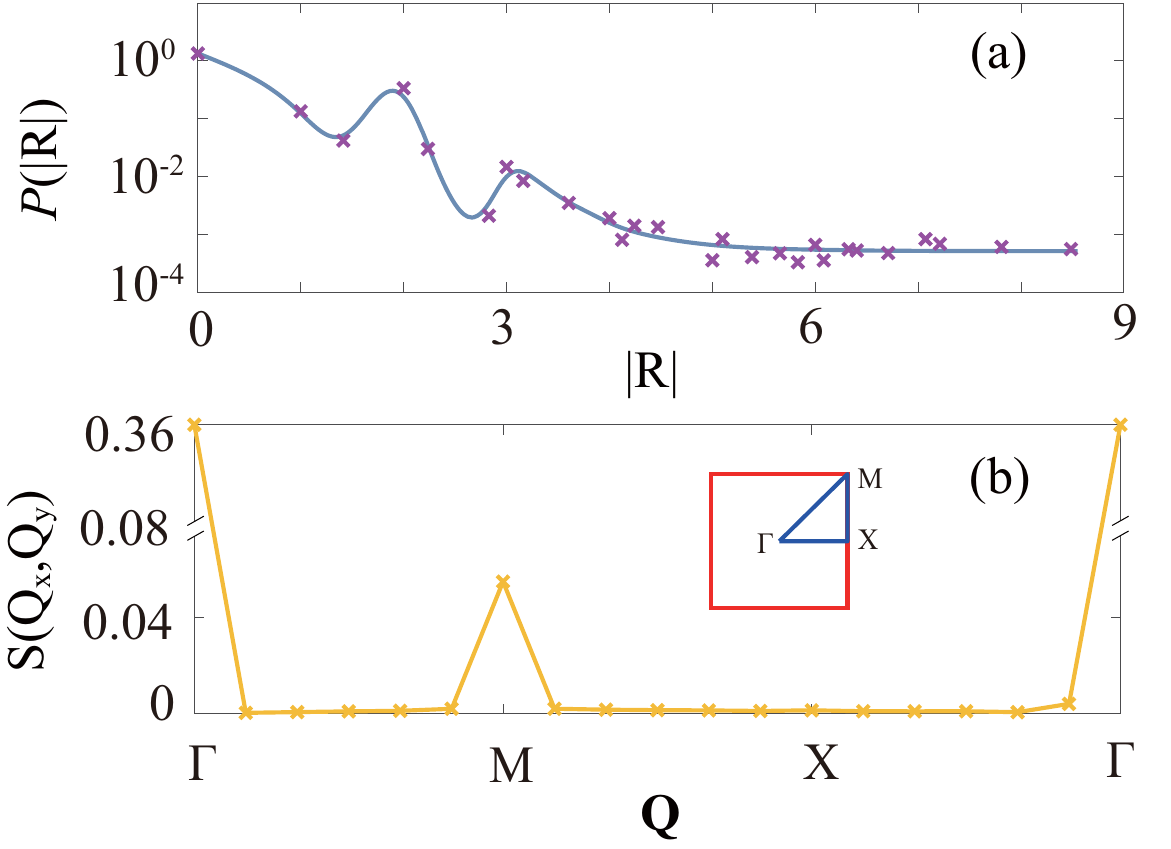}
\end{center}
\caption{(a) Pairing correlation $P(|\mathbf{R}|)$ as a
function of $|\mathbf{R}|$. $P(|\mathbf{R}|)$ shows saturated long-ranged
correlation, indicating the existence of superfluid pairing order. (b) Structure factor $S(\mathbf{Q})$ as a function of the momentum, where its peak is located
at $(\pi/a,\pi/a)$, indicating that there is a checkerboard density pattern. Here $J=0.14$, $n_{0}=0.6$ and other parameters are the same as in
Fig.~\ref{fig:1}.}
\label{fig:3}
\end{figure}

To further verify the existence of CDW and superfluid orders, we have
performed a variational Monte Carlo (VMC)
calculation on a $12\times 12$ lattice system with periodic boundary
condition~\cite{MC1_PhysRevB,MC2_GROS}. Regarding the superfluid order in the ground state, we
study the pairing correlation through the VMC method. For instance, considering the dominant pairing component
$\Delta [\sin(2k_{x}a) + i \sin(2k_{y}a)]$, the correlation can be defined as
\begin{equation}
P(\mathbf{R})=\frac{1}{2N_{L}}\sum\limits_{\mathbf{R}_{i}}\langle
\Delta ^{\dagger }(\mathbf{R}_{i})\Delta (\mathbf{R}_{i}+\mathbf{R})+\Delta (%
\mathbf{R}_{i})\Delta ^{\dagger }(\mathbf{R}_{i}+\mathbf{R})\rangle ,
\label{4}
\end{equation}%
with $\Delta (\mathbf{R}_{i})\equiv
c_{i}c_{i+2e_{x}}-c_{i}c_{i-2e_{x}}+i(c_{i}c_{i+2e_{y}}-c_{i}c_{i-2e_{y}})$.
$\mathbf{R}$ is an 2D vector in the $xy$-plane. As
shown in Fig.~\ref{fig:3}(a), the long-ranged saturation behavior of the
pairing correlation $P(\mathbf{R})$ indicates the
existence of superfluid pairing order in the ground state. While to verify the
existence of CDW , we calculate the density
structure factor defined as
\begin{equation}
S(\mathbf{Q})=\frac{1}{N_{L}^{2}}\sum\limits_{i,j}
\langle c_{i}^{\dagger }c_{i}c_{j}^{\dagger }c_{j}\rangle e^{i\mathbf{Q\cdot
(R_{i}-R_{j})}}.  \label{5}
\end{equation}%
The peak in density structure factor provides information on the CDW order.
As shown in Fig.~\ref{fig:3}(b), when J beyond the threshold, the structure factor $S(\mathbf{Q})$ is peaked
at $(\pi /a,\pi /a)$, indicating the existence of a checkerboard density
pattern  in the ground state, which is consistent with our  mean-field  calculations as shown in Fig.~\ref{fig:1}.

\textit{Chiral odd Chern number lattice supersolids $\raisebox{0.01mm}{---}$} In the following, we will study the topological nature of the SS phase.
As shown in Fig.~\ref{fig:1}, there are three topologically distinct SS phases.
One topological trivial region and two topologically  non-trivial regions can be distinguished by the Chern number $C =\frac{i}{2\pi }\sum_{E_{n}<0}\int_{{}}dk_{x}dk_{y}(\langle \partial
_{k_{y}}\phi _{n}({k})|\partial _{k_{x}}\phi _{n}({k})\rangle
-\langle \partial _{k_{x}}\phi _{n}({k})|\partial _{k_{y}}\phi _{n}({k}%
)\rangle )$, where $\phi _{n}({k})$ is the eigenstate with energy $E_{n}$ of Eq. (1) under the mean-field approximation. We find that the topological trivial
region SS-I phase is characterized with the zero Chern number. While the two topological regions, SS-II and SS-III, are featured by the non-zero Chern number. More interestingly, we find that both  SS-III and SS-II are characterized with an odd Chern number, i.e., $C=1$ and $C=3$, respectively, which can support unpaired CMFM, to be shown below.

\begin{figure}[t]
\begin{center}
\includegraphics[width=8cm]{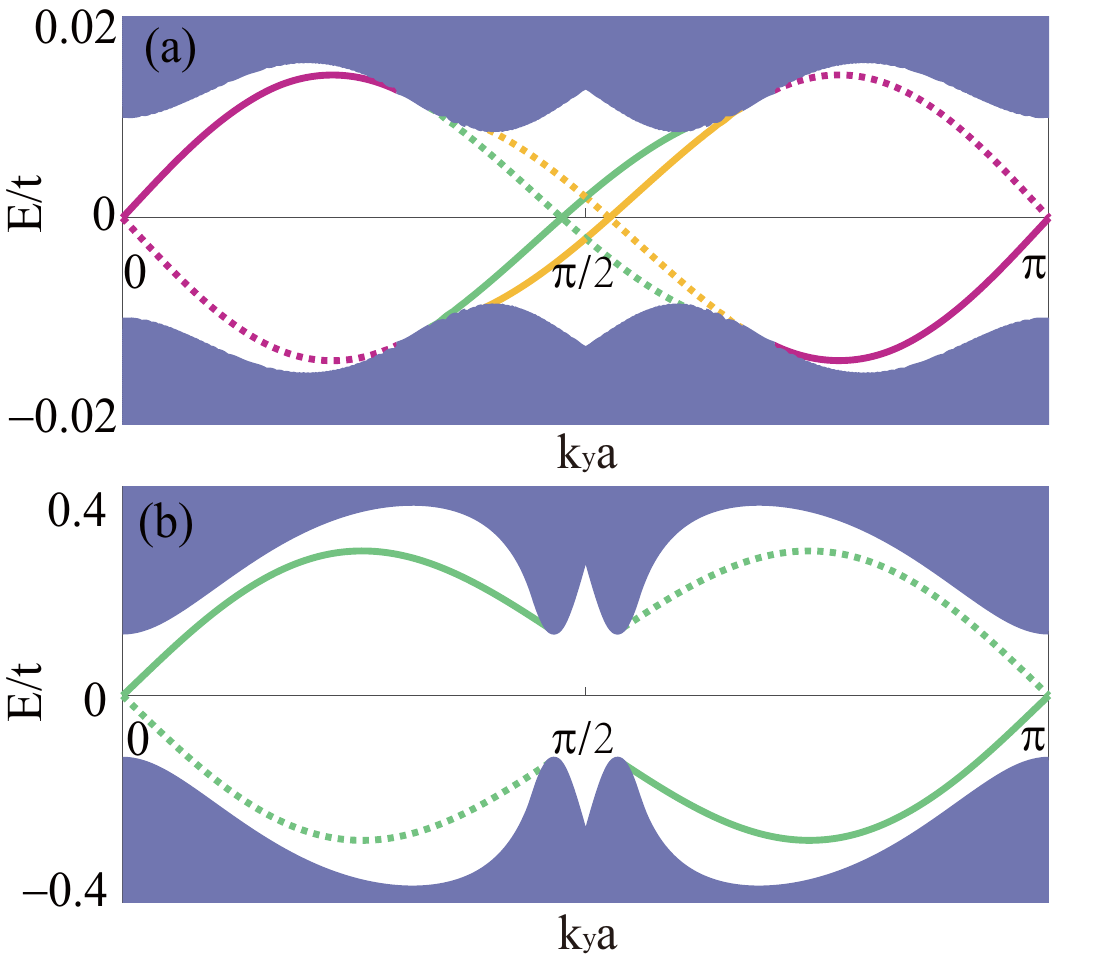}
\end{center}
\caption{(a) and (b) Energy spectrum of the system with open (periodic) boundary
conditions in the $x$ ($y$) directions. In (a), there are three pairs of chiral edge modes in SS-II, where $J=0.27$, $n_{0}=0.64$ . In (b), there is one pair of chiral edge modes in phase SS-III, where $J=0.53$, $n_{0}=0.82$. Other parameters are the same as in Fig.~\ref{fig:1}.}%
\label{fig:4}%
\end{figure}

\begin{figure}[t]
\begin{center}
\includegraphics[width=8.5cm]{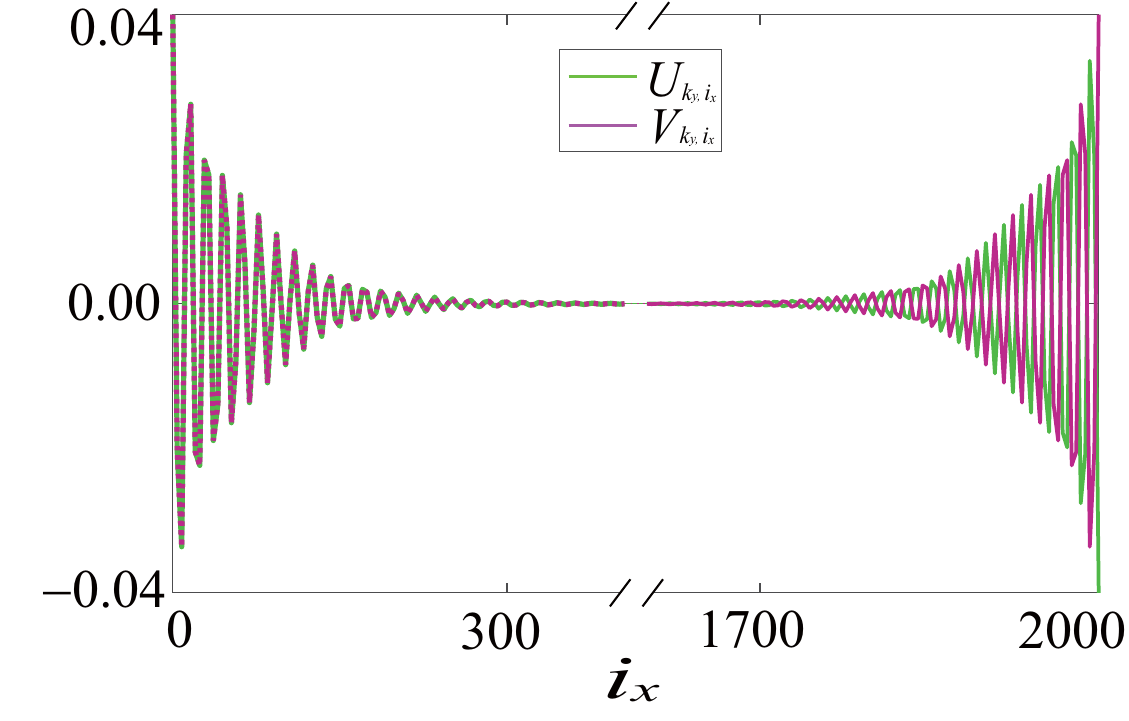}
\end{center}
\caption{The wavefunction of the zero-energy state in Fig.~\ref{fig:4} (a). Here, we choose $k_ya=0$. It turns out that the zero-energy edge state is a chiral Majorana fermion mode.}%
\label{fig:5}%
\end{figure}

To gain more insight into the topological property of the system, we have applied a series of unitary transformations (see details in SM) to reform the BdG Hamiltonian
in a much clearer way as
\begin{equation}
\mathscr{H}_{BdG}\equiv \left(
\begin{array}{cc}
H_{SF}^{^{\prime }} & 0_{2\times 2} \\
0_{2\times 2} & H_{SF}^{\prime \prime }%
\end{array}%
\right),  \label{10}
\end{equation}%
with $
H_{SF}^{^{\prime }}=\left(
\begin{array}{cc}
\sqrt{{\xi _{\mathbf{k}}^{2}+\delta }^{2}}{-\mu } & \Delta (\mathbf{k}) \\
\Delta ^{\ast }(\mathbf{k}) & -\sqrt{{\xi _{\mathbf{k}}^{2}+\delta }^{2}}+{%
\mu }%
\end{array}%
\right)$ and $H_{SF}^{\prime \prime }=\left(
\begin{array}{cc}
-\sqrt{{\xi _{\mathbf{k}}^{2}+\delta }^{2}}{-\mu } & \Delta (\mathbf{k}) \\
\Delta ^{\ast }(\mathbf{k}) & \sqrt{{\xi _{\mathbf{k}}^{2}+\delta }^{2}}+{%
\mu }%
\end{array}%
\right)$.
Here, to simplify the analysis, we take the dominant component of superfluid order and $\Delta(\mathbf{k})$ in Eq.~\eqref{10} is approximated as $\Delta [\sin(2k_{x}a) + i \sin(2k_{y}a)]$.
Then, from Eq.~\eqref{10}, we can understand the topology of the system. First, we find that $H_{SF}^{\prime \prime}$ is always typologically trivial when considering $\mu>0$,  which is identified by the vanished Chern number.
Second, we also find that there are three distinct topological regions for $H^{\prime}$: (i) ${\mu >}\sqrt{{\delta }^{2}+{16t}^{2}}$ or ${0<\mu <\delta }$, $%
H_{SF}{^{\prime }}$ are engineered in the topological trival region with
Chern number $C=0$; (ii) $\sqrt{{\delta }^{2}+{4t}^{2}}{<{\mu <}}\sqrt{{\delta }^{2}+{16t}^{2}}$%
, $H_{SF}^{^{\prime }}$ are tuned in topological regions with Chern number $%
C=1$; (iii) ${\delta <{\mu <}}\sqrt{{\delta }^{2}+{4t}^{2}}$, a topological
phase characterized by the Chern number $C=3$ is achieved.
Therefore,
distinct topological regions of $\mathscr{H}_{BdG}$ can be engineered by tuning the CDW order. Amazingly, such a scheme can be naturally achieved in our proposed SS phase when varying the interaction strength and the average filling.  As shown in Fig.~\ref{fig:1}, in the SS region, there are three distinct topological regions. When the system is close to the half filling, the SS phase is in the topological trivial region, i.e., SS-I phase. When further
increasing the average filling, there are two topological phase transitions and the SS phase enters two  distinct topological non-trivial regions with two different Chern numbers $C=3$ and $C=1$, respectively. Therefore, an unveiled
CLSS phase is achieved .

\textit{Multiple number-tunable chiral Majorana fermions $\raisebox{0.01mm}{---}$} Since the Chern number counts the number of CMFM at the edge of the system,  an odd Chern number corresponds to the unpaired chiral Majorana edge mode~\cite{Kitaev_2001}, which constitutes a non-Abelian phase of matter. Therefore, our proposed CLSS phases with distinct odd Chern numbers can support the number-tunable unpaired chiral Majorana edge modes. To show that
a cylinder geometry is chosen in the $xy$-plane, i.e., considering the open
(periodic) boundary conditions along the $x(y)$ directions, respectively.
The edge excitations can be obtained (see details in SM). For instance,  as shown in Fig.~\ref{fig:4}(a), for SS-II phase,  it is shown that all the bulk modes are gapped and there are three pairs of chiral edge states located at two outer edges of the system, because the Chern number of SS-II phase is $C=3$ satisfying the so-called
bulk-edge correspondence.  More interestingly, we also find that among these chiral edge modes there are six zero-energy edge states.  The wavefunction of that can be expressed as $(u_{k_{y},i_{x}}^{0},v_{k_{y},i_{x}}^{0},u_{k_{y}^{\prime
},i_{x}}^{0},v_{k_{y}^{\prime },i_{x}}^{0})=(U_{k_{y}\mathbf{,}%
i_{x}}e^{i\theta _{k_{y}\mathbf{,}i_{x}}},V_{k_{y},i_{x}}e^{-i\theta _{k_{y}%
\mathbf{,}i_{x}}},U_{k_{y}^{\prime },i_{x}}e^{i\theta _{k_{y}^{\prime }%
\mathbf{,}i_{x}}},V_{k_{y}^{\prime },i_{x}}e^{-i\theta _{k_{y}^{\prime }%
\mathbf{,}i_{x}}})$, which satisfies $u_{k_{y}(k_{y}^{\prime})\mathbf{,}%
i_{x}}^{0}=v_{k_{y}(k_{y}^{\prime}),i_{x}}^{0*}$ on the left edge and $u_{k_{y}(k_{y}^{\prime}),i_{x}}^{0}=-v_{k_{y}(k_{y}^{\prime}),i_{x}}^{0*}$
on the right edge, for instance, as shown in Fig.~\ref{fig:5}. Therefore, these six  zero-energy eigenstates support three unpaired  chiral Majorana fermions localized at each edge of the system. While for SS-III phase, as shown in Fig.~\ref{fig:4}(b), since the Chern number $C=1$, it is found that there are two zero-energy eigenstates which support one unpaired chiral Majorana fermion at each edge of the  system. Therefore, the number-tunable unpaired chiral Majorana edge modes can be achieved in our proposed CLSS phase, which would offer an intriguing possibility pointing to braiding statistics and applications to topological quantum computing.

\textit{Conclusion $\raisebox{0.01mm}{---}$} We find a new type of topological lattice supersolid state of a single component Rydberg-dressed Fermi gas in an optical lattice,
which arises from the unveiled effect induced by the competition between
two distinct length scales, i.e., lattice period and the distance of resonant
Rydberg-dressing. Such a scheme thus drummed up a new way of engineering
RDI and new types of many-body phases can be achieved, which should be
observable in future experiments.

\textit{Acknowledgment $\raisebox{0.01mm}{---}$} This work is supported by the National Key R$\&$D Program of China (2021YFA1401700), NSFC (Grants No. 12074305, 12147137, 11774282), the National Key Research and Development Program of China (2018YFA0307600), Xiaomi Young Scholar Program. We also thank the HPC platform of Xi'An Jiaotong University, where our numerical calculations was performed.

\bibliographystyle{apsrev}
\bibliography{SS4}

\onecolumngrid

\renewcommand{\thesection}{S-\arabic{section}}
\setcounter{section}{0}  
\renewcommand{\theequation}{S\arabic{equation}}
\setcounter{equation}{0}  
\renewcommand{\thefigure}{S\arabic{figure}}
\setcounter{figure}{0}  

\indent

\begin{center}\large
\textbf{Supplementary Material:\\ Chiral odd Chern number lattice supersolidity with tunable unpaired Majorana fermions in a Rydberg-dressed Fermi gas}
\end{center}

\section{Mean-field method}

In this section, we will provide more details about the mean-field method.
Under the mean-field approximation, the Hamiltonian in Eq.(1) of the main text can be rewritten in the momentum space as
\begin{eqnarray}
\mathbf{H}_{MF} &=&\sum_{\mathbf{k}}\xi _{\mathbf{k}}c_{\mathbf{k}}^{\dagger
}c_{\mathbf{k}}+\sum_{\mathbf{k}}({\frac{\Delta ({\mathbf{k}})}{2}}c_{%
\mathbf{k}}^{\dagger }c_{-\mathbf{k}}^{\dagger }+h.c.)
+\sum_{\mathbf{Q_{m}}=\mathbf{\pm Q}}\sum_{\mathbf{k}}({\frac{{\delta_{\mathbf{Q_m}} }}{4}}c_{\mathbf{k}}^{\dagger }c_{%
\mathbf{k+Q_m}}+h.c.)-E_{I},.  \label{3}
\end{eqnarray}%
where $\xi _{\mathbf{k}}=\varepsilon _{\mathbf{k}}-\mu $ and  $E_{I}=\frac{1}{2}\sum_{i\neq j}V_{i-j}(-n_{i}n_{j}+|\langle
c_{j}c_{i}\rangle |^{2})$. Here, $\varepsilon _{\mathbf{k}}=-2t(\cos {k_{x}a}+\cos {k_{y}a}%
)$ is the band dispersion and $\mu $ is the chemical potential.
Through diagonalizing Eq. (S1) via the Bogoliubov method,  we can obtain the mean-field ground state energy of the system as $E_{MF}=\sum\limits_{n}E_{n}\Theta (-E_{n})+\frac{1}{2}\sum_{%
\mathbf{k}}\xi _{\mathbf{k}}-E_{I}$, where $E_{n}$ labels the $n$-th eigenenergy of Eq. (S1) and $\Theta$ is the Heaviside step function.
 The order parameters defined in Eq. (S1) can be obtained by minimizing $E
_{MF}$ for a certain average filling of the system determined by the
relation $n_{0}=-\frac{1}{N_{L}}\frac{\partial {E_{MF}}}{{\partial {%
\mu }}}$.

\section{ variational Monte Carlo method}
In this section, we will provide a detailed description
of the variational Monte Carlo (VMC) method used in this work.
The VMC method is one of promising methods to study strongly
correlated systems ~\cite{MC1_PhysRevB,MC2_GROS} and there is no sign problem in studies of fermionic systems since the weight of Monte Carlo sampling is positive definite. The wave function employed in our many-variable variational Monte Carlo (mVMC) simulation can be expressed as $\left\vert \phi _{\mathrm{ref}%
}\right\rangle =\mathcal{P}_{J}\left\vert \phi _{\mathrm{pair}}\right\rangle
$, where $\left\vert \phi _{\mathrm{pair}}\right\rangle
=[\sum\limits_{i,j=1}^{N_{L}}f_{ij}c_{i}^{\dagger }c_{j}^{\dagger
}]^{N/2}\left\vert 0\right\rangle $ is the Pfaffian pairing wave function and $\mathcal{P}_{\mathrm{J}}=\exp \left[ \frac{1}{2%
}\sum_{i\neq j}v_{ij}\left( n_{i}-1\right) \left( n_{j}-1\right) \right] $
is the Jastrow factor, which accounts for long-ranged
density correlations. Here $N$ refers to
the number of fermions. Such a flexible variational wavefunction with a
large number of variational parameters can be simultaneously optimized by
using the stochastic reconfiguration (SR) method, which
can be applied to efficiently compute the ground state of our proposed
system.

\section{The topological nature of the system}
The topological nature of the system can be understood through the Bogliubov-de Gennes
(BdG) Hamiltonian
\begin{equation}
H_{BdG}=\left(
\begin{array}{cccc}
{\xi _{\mathbf{k}}} & \Delta (\mathbf{k}) & {\delta } & 0 \\
\Delta ^{\ast }(\mathbf{k}) & -{\xi _{-\mathbf{k}}} & 0 & -{\delta } \\
{\delta } & 0 & {\xi _{\mathbf{k}+\mathbf{Q}}} & \Delta (\mathbf{k}+\mathbf{Q%
}) \\
0 & -{\delta } & \Delta ^{\ast }(\mathbf{k}+\mathbf{Q}) & -{\xi _{-\mathbf{k}%
-\mathbf{Q}}}%
\end{array}%
\right)  \label{7}
\end{equation}%
where the Nambu spinors are chosen as $(c_{\mathbf{k}}^{\dag },c_{-\mathbf{k}%
},c_{\mathbf{k}+\mathbf{Q}}^{\dag },c_{-\mathbf{k}-\mathbf{Q}})$.
To simplify the analysis, here we take the dominant component of superfluid order and $\Delta(\mathbf{k})$ is approximated as $\Delta [\sin(2k_{x}a) + i \sin(2k_{y}a)]$.
Then, we apply a series of unitary transformations to the BdG Hamiltonian in Eq. (S2) and we obtain%
\begin{eqnarray}
\mathscr{H}_{BdG} &=&\Lambda ^{\dagger }H_{BdG}\Lambda  \notag \\
&=&\left(
\begin{array}{cccc}
{E}_{1}^{\pi } & \Delta (\mathbf{k}) & {0} & 0 \\
\Delta ^{\ast }(\mathbf{k}) & -{E}_{1}^{\pi } & {0} & 0 \\
{0} & {0} & {E}_{2}^{\pi } & \Delta (\mathbf{k}) \\
0 & 0 & \Delta ^{\ast }(\mathbf{k}) & -{E}_{2}^{\pi }%
\end{array}%
\right)  \label{8}
\end{eqnarray}%
where $\Lambda =T^{-1}\Gamma T$, with
$T=\left(
\begin{array}{cccc}
{1} & 0 & 0 & 0 \\
0 & {0} & 1 & 0 \\
0 & 1 & {0} & 0 \\
0 & 0 & 0 & 1%
\end{array}%
\right)$ and $\Gamma =\left(
\begin{array}{cc}
\Gamma _{CDW} & 0_{2\times 2} \\
0_{2\times 2} & \Gamma _{CDW}%
\end{array}%
\right) $. Here, $\Gamma _{CDW}$ can be constructed through the relation $\Gamma
_{CDW}^{\dagger }\left(
\begin{array}{cc}
{\xi _{\mathbf{k}}} & {\delta } \\
{\delta } & {\xi _{\mathbf{k}+\mathbf{Q}}}%
\end{array}%
\right) \Gamma _{CDW}=\left(
\begin{array}{cc}
{E}_{1}(\mathbf{k}) & {0} \\
{0} & {E}_{2}(\mathbf{k})%
\end{array}%
\right) $ with $E_{1}(\mathbf{k})=\sqrt{{\xi _{\mathbf{k}}^{2}+\delta }^{2}}{%
-\mu }$ and $E_{2}(\mathbf{k})=-\sqrt{{\xi _{\mathbf{k}}^{2}+\delta }^{2}}{%
-\mu }$.$\ ${Then, }$\mathscr{H}_{BdG}${\ can be rewritten}, i.e.,
$\mathscr{H}_{BdG}\equiv \left(
\begin{array}{cc}
H_{SF}^{^{\prime }} & 0_{2\times 2} \\
0_{2\times 2} & H_{SF}^{\prime \prime }%
\end{array}%
\right)$, as shown in the main text.
\section{Edge excitations}
To show the edge excitations of our proposed CLSS phase, we consider a cylinder
geometry in the $xy$-plane, i.e., choosing the open (periodic) boundary conditions
along the x(y) directions, respectively. Then, the edge excitations can be obtained  through solving the following eigen-problem
\begin{equation}
\sum\limits_{j_{x}}\left(
\begin{array}{cccc}
H_{i_{x},j_{x}}(k_{y}) & \Delta _{i_{x},j_{x}}(k_{y}) & \delta \delta
_{i_{x},j_{x}} & 0 \\
\Delta _{i_{x},j_{x}}^{\ast }(k_{y}) & -H_{i_{x},j_{x}}(k_{y}) & 0 & -\delta
\delta _{i_{x},j_{x}} \\
\delta \delta _{i_{x},j_{x}} & 0 & H_{i_{x},j_{x}}(k_{y}^{\prime }) & \Delta
_{i_{x},j_{x}}(k_{y}^{\prime }) \\
0 & -\delta \delta _{i_{x},j_{x}} & \Delta _{i_{x},j_{x}}^{\ast
}(k_{y}^{\prime }) & -H_{i_{x},j_{x}}(k_{y}^{\prime })%
\end{array}%
\right) \left(
\begin{array}{c}
u_{k_{y},j_{x}}^{n} \\
v_{k_{y},j_{x}}^{n} \\
u_{k_{y}^{\prime } ,j_{x}}^{n} \\
v_{k_{y}^{\prime } ,j_{x}}^{n}%
\end{array}%
\right) =E_{n}\left(
\begin{array}{c}
u_{k_{y},i_{x}}^{n} \\
v_{k_{y},i_{x}}^{n} \\
u_{k_{y}^{\prime },i_{x}}^{n} \\
v_{k_{y}^{\prime },i_{x}}^{n}%
\end{array}%
\right) ,
\label{13}
\end{equation}%
where the momentum  $%
k_{y}^{\prime }=k_{y}+\pi/a $. $H_{i_{x},j_{x}}(k_{y})=-t(%
\delta _{i_{x}+1,j_{x}}+\delta _{i_{x},j_{x}+1})+(-2t\cos k_{y}a-\mu )\delta
_{i_{x},j_{x}}$, $\Delta _{i_{x},j_{x}}(k_{y})=\sum_{m,n}\frac{\Delta _{m,n}}{2i}(e^{i{na}k_{y}}\delta _{i_{x}+m,j_{x}}-e^{-i{na}k_{y}}\delta _{i_{x},j_{x}+m})$.

\end{document}